\documentclass{kluwer}    % Specifies the document style.
\usepackage{epsfig}
\usepackage{graphicx}

\begin{document}
\begin{article}
\begin{opening}
\title{Unidentified $\gamma$-ray sources off the Galactic plane as low-mass microquasars?}
\author{I.~A. \surname{Grenier}}
\institute{Universit\'{e} Paris VII \& Service d'Astrophysique, CEA Saclay, 91191 Gif/Yvette, France}
\author{M.~M. \surname{Kaufman Bernad\'{o}} \footnote{Fellow of CONICET}}
\author{G.~E. \surname{Romero}\footnote{Member of CONICET}}
\institute{Instituto Argentino de Radioastronom\'{i}a, C.C. 5, 1894 Villa Elisa, Argentina}
\date{Received 2004 August 8; accepted 2004 ...}

\begin{abstract}
A subset of the unidentified EGRET $\gamma$-ray sources with no active galactic nucleus or other conspicuous
counterpart appears to be concentrated at medium latitudes. Their long-term variability and their spatial
distribution indicate that they are distinct from the more persistent sources associated with the nearby
Gould Belt. They exhibit a large scale height of $1.3 \pm 0.6$ kpc above the Galactic plane. Potential
counterparts for these sources include microquasars accreting from a low-mass star and spewing a continuous
jet. Detailed calculations have been performed of the jet inverse Compton emission in the radiation fields
from the star, the accretion disc, and a hot corona. Different jet Lorentz factors, powers, and aspect angles
have been explored. The up-scattered emission from the corona predominates below 100 MeV whereas the disc and
stellar contributions are preponderant at higher energies for moderate ($\sim 15^{\circ}$) and small ($\sim
1^{\circ}$) aspect angles, respectively. Yet, unlike in the high-mass, brighter versions of these systems,
the external Compton emission largely fails to produce the luminosities required for 5 to 10 kpc distant
EGRET sources. Synchrotron-self-Compton emission appears as a promising alternative.
\end{abstract}
\end{opening}

\section{Variable $\gamma$-ray sources in the Galactic disc}
The EGRET telescope has detected 263 $\gamma$-ray sources above 100 MeV \cite{3eg}, half of which have been
firmly or plausibly associated with flat radio spectrum AGN and a handful of nearby radiogalaxies
\cite{sowards03,sowards04,mattox01}. Six are identified with pulsars. The other 126 sources remain
unidentified because of their poor localization, typically within a degree, or their faintness at lower
energies.

Thirty three (33) bright sources along the Galactic plane are associated with star-forming regions that harbour many likely
$\gamma$-ray emitters, steady ones such as pulsars and supernova remnants, and more variable ones such as
pulsar wind nebulae, massive star binaries, and accreting X-ray binaries \cite{romero99,grenier04}.
\inlinecite{nolan03} found a group of 17 sources, concentrated in the inner Galaxy, that exhibit variability
on timescales of weeks to months, hence ruling out a supernova remnant or a young pulsar origin. This group
is brighter than the other low-latitude sources \cite{bosch04}. Their luminosity ranges from 0.8 to $20 \times 10^{34}$ $(D/\;{5\;\rm kpc})^2$ erg s$^{-1}$ sr$^{-1}$ above 100 MeV. They may be associated with variable
pulsar wind nebulae or with lower-energy versions of the colliding winds from a pulsar and a massive star, as
recently observed at TeV energies by HESS in PSR B1259-63 \cite{kirk99}.

\inlinecite{kaufman02} have alternatively suggested that they are microquasars with persistent jets and
high-mass stellar companions. The long-term variability may be caused by the precession of the accretion
disc, henceforth of the jet, that is induced by the gravitational torque of the star, or by variations in the
accretion rate along an eccentric orbit, or by instabilities and inhomogeneities in the stellar wind.
Synchro-self-Compton (SSC) radiation in the expanding radio blobs ejected during sporadic flares shine too
briefly to account for EGRET sources \cite{atoyan99}. Three microquasars with high-mass companions indeed
coincide with EGRET sources: LS5039 in 3EG J1824-1514, LSI +61 303 in 3EG J0241+6103, and AX J1639.0-4642 in
3EG J1639-4702 (see Rib\'o et al, these proceedings). The former and the latter do not exhibit variability in
the EGRET data \cite{nolan03}. The jet from LSI +61 303 is known to precess and the observed $\gamma$-ray
flux is variable, though it does not correlate with the radio flares (\inlinecite{kniffen97}; see , however, \inlinecite{massi04}). Energetic electrons or
pairs in the jet can spawn $\gamma$-rays by up-scattering photons from the bright UV star, the soft X-ray
accretion disc, a hard X-ray corona, or their own synchrotron radiation. External inverse Compton (EC)
emission from a cylindrical jet \cite{romero04}, as well as SSC emission or EC emission from an expanding jet
\cite{bosch04}, can both reproduce the observed luminosities. SSC emission predominates for large jet
magnetic fields, typically $>$ 100 G. EC emission yields a large variety of $\gamma$-ray spectra, depending
on the relative contributions of the different radiation fields and on the jet/accretion power ratio. The
latter should typically exceed $10^{-4}$ to power an EGRET source.

%----------------------------------------------------------------
\begin{figure}
\centerline{\includegraphics[width=4in]{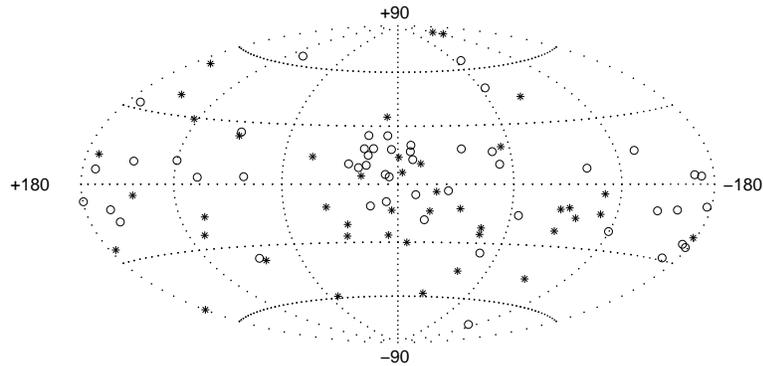}} \caption{All-sky plot, in Galactic coordinates, of
the unidentified EGRET sources at latitudes $|b|>3^{\circ}$. The steadier sources associated with the Gould
Belt and the variable sources are marked as circles and stars, respectively.} \label{fig:mapufos}
\end{figure}
%----------------------------------------------------------------
%----------------------------------------------------------------
\begin{figure}
\centerline{\includegraphics[width=4in]{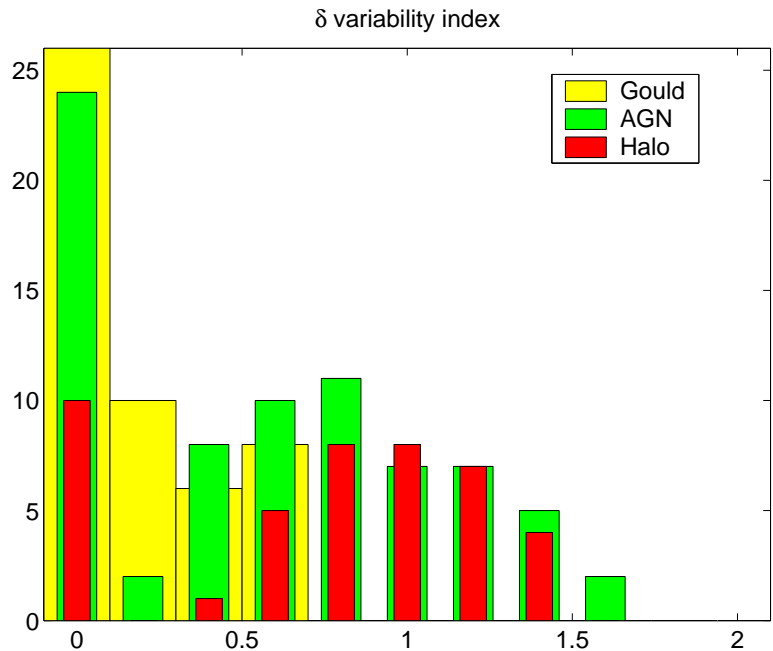}} \caption{Distributions of the $\delta$ variability
indices from Nolan et al. (2003) for the firm AGN, Gould Belt, and halo sources at latitudes
$|b|>3^{\circ}$.} \label{fig:deltahist}
\end{figure}
%----------------------------------------------------------------

\section{Variable $\gamma$-ray sources off the Galactic plane}
The 93 unidentified sources detected away from the Galactic plane are displayed in Figure \ref{fig:mapufos}.
Their concentration at $3^{\circ} < |b| < 30^{\circ}$ and in the inner half steradian clearly indicates (at a
7$\sigma$ confidence level) that 70 to 100 \% of them have a Galactic origin, depending on the choice of
Galactic scale height. Their temporal and spatial characteristics reveal a heterogeneous sample. A subset of
steadier sources, marked as circles in Figure \ref{fig:mapufos}, has been associated with the nearby Gould
Belt \cite{gehrels00,grenier00}. The others are clearly more variable. Figure \ref{fig:deltahist} shows their
variability index distribution which closely follows that of the variable firm AGN sources \cite{nolan03}.
The two distributions are consistent under the Kolmogorov (K) test or the T test developed by
\inlinecite{eadie}, and both are significantly at variance with the Belt source distribution (5$\sigma$ and
4$\sigma$). The average $\overline{\delta}$ indices of the three sets are 0.79 $\pm$ 0.08, 0.66 $\pm$ 0.06,
and 0.42 $\pm$ 0.06 for the variable, AGN, and Belt sources, respectively. The spectral index distributions
of the 3 sets are quite consistent according to the K and T tests, so the slightly softer average index of
2.43 $\pm$ 0.05 for the variable sources, compared with 2.23 $\pm$ 0.02 for the Belt and AGN sources, is not
significant. Taking into account the detection biases \cite{grenier00}, the spatial distribution of the
variable sources implies an origin in a thick Galactic disk with a scale height of 1.3 $\pm$ 0.6 kpc. A fit
to the whole, persistent and variable, source sample with various combinations of Galactic distributions
nearly equally shares the sources between the Gould Belt and a large scale height component, with a
3.4$\sigma$ and a 4.2$\sigma$ improvement over the single distribution fits. At typical distances of 5 to 10
kpc, the luminosities of the variable sources range from 2 to $30 \times 10^{34}$ erg s$^{-1}$ sr$^{-1}$ above 100
MeV. They exhibit large luminosity $L_{\gamma}/L_X$ ratios of a few hundred.

The compact objects likely to power sources high above the plane include ms pulsars and microquasars with a
low-mass star companion, both having migrated away from the Galactic plane or escaped from globular clusters.
None of the $\gamma$-ray sources coincides with a globular cluster. Pulsed $\gamma$-rays have been detected
from the ms pulsar PSR J0218+4232, in phase with the radio and X-ray peaks \cite{kuiper02}. This object
shares many traits with the halo sources: a distance of 5.7 kpc and an altitude of 1.6 kpc, a luminosity of
$1.6 \times 10^{34}$ erg s$^{-1}$ sr$^{-1}$ and a spectral index of 2.6 above 100 MeV. PSR J0218+4232 does not
belong to a globular cluster either. Yet, no long-term variability is expected from theory
\cite{zhangcheng03}. In the next section, we explore whether low-mass microquasars can produce the halo
sources despite their intrinsic faintness and softness compared with the young high-mass systems. Their
stellar and disc intensities are reduced by 4-6 and 1 order of magnitude, respectively, and their thermal
emissions peak a decade or two lower in energy.

\section{Emission from low-mass microquasars}
Microquasars are found up to very high Galactic latitude and height above the plane \cite{mirabel01}. For
instance, XTE J1118+480, at $b = 62^{\circ}$ and a distance of $1.8 \pm 0.6$ kpc, lies at 1.6 $\pm$ 0.5 kpc
above the plane \cite{mcclintock01a}. The central object accretes matter through a disc from a low-mass star
via Roche lobe overflow. \inlinecite{mcclintock01a} and \inlinecite{wagner01} constrained the companion
spectral type to be between K5 V and M1 V. A large mass function, $f(M) \approx$ 6 M$\odot$, strongly
suggests that the compact object is a black hole in a fairly compact binary system with a short orbital of
4.1 hr \cite{wagner01}. The disc may be precessing under the stellar tidal influence \cite{torres02_XTE1118}.
The rapid and correlated UV-optical-X-ray variability in the low-hard state is interpreted as a signature of
the strong coupling between a hot corona and a jet emitting synchrotron radiation up to, at least, the UV
band \cite{hynes03,chaty03,malzac04}. So, pairs in the jet can reach 10 GeV in a modest magnetic field of 10
G. Whether the synchrotron emission extends to hard X rays is quite possible since the data requires the jet
to dominate the energy budget and drive the variability, so jet pairs could reach hundreds of GeV. The
outflow has remained steady through the outburst evolution \cite{chaty03}. The coronal emission extends to
$\sim 150$ keV \cite{mcclintock01b}. The optical to hard-X-ray data can me modelled by the Comptonisation in
a hot corona or in the inner accretion flow of the soft photons emitted by the outer cold disc with an inner
radius of $\sim 55 R_{\rm Schw}$ and a temperature of $\sim 24$ eV \cite{esin01,mcclintock01b,malzac04}.
%----------------------------------------------------------------
\begin{table} %
\begin{tabular}{ll}
\hline
black hole mass           & $M_{\rm bh} = 6.5$ M$\odot$\\
mass accretion rate       & $\dot{M} = 3 \times 10^{-8}$ M$\odot$ yr$^{-1}$\\
K-M star bolometric luminosity & $L_{\rm KM} = 4 \times 10^{32}$ erg/s\\
F star bolometric luminosity & $L_{\rm F} = 1.5 \times 10^{34}$ erg/s\\
star temperature          & $kT_{\rm KM} = 1$ eV and $kT_{F} = 1.8$ eV\\
star orbital radius       & $D_* = 1.7 \times 10^{11}$ cm\\
jet/accretion power ratio & $q_{\rm jet} = P_{\rm jet}/\dot{M}c^2 = 10^{-3}$ to $10^{-2}$\\
corona luminosity         & $L_{\rm cor} = 7.8 \times 10^{34}$ erg/s\\
corona outer radius       & $R_{\rm cor} = \, 10^8$ cm\\
corona photon index ($dN_X/dE \propto E_X^{-\alpha}$) & $\alpha_{\rm cor} = 1.8$\\
corona cut-off energy     & $E_{\rm cor} = 150$ keV\\
disc luminosity           & $L_{\rm disc} = 8.6 \times 10^{35}$ erg/s\\
disc temperature          & $kT_{\rm disc} = 24$ eV \\
initial jet radius        & $R_{\rm jet} = 1.9 \times 10^7$ cm\\
jet bulk Lorentz factor   & $\Gamma_{\rm jet} = 3$ to $10$\\
jet viewing angle         & $\Phi = 1^{\circ}$ to $30^{\circ}$\\
jet electron index ($dN_e/dE \propto E_e^{-p}$) & $p = 2$ to 3\\
maximum electron energy   & $E_{e\,\rm max} = 5$ GeV to 5 TeV\\
minimum electron energy   & $E_{e\,\rm min} = 1$ to 5 MeV\\
\hline
\end{tabular}
\caption[]{Parameter set used in the model}\label{parset}
\end{table}
%----------------------------------------------------------------

\begin{figure}%[H]
%\tabcapfont
\centerline{%
\begin{tabular}{l@{\hspace{1pc}}r}
\includegraphics[width=4in]{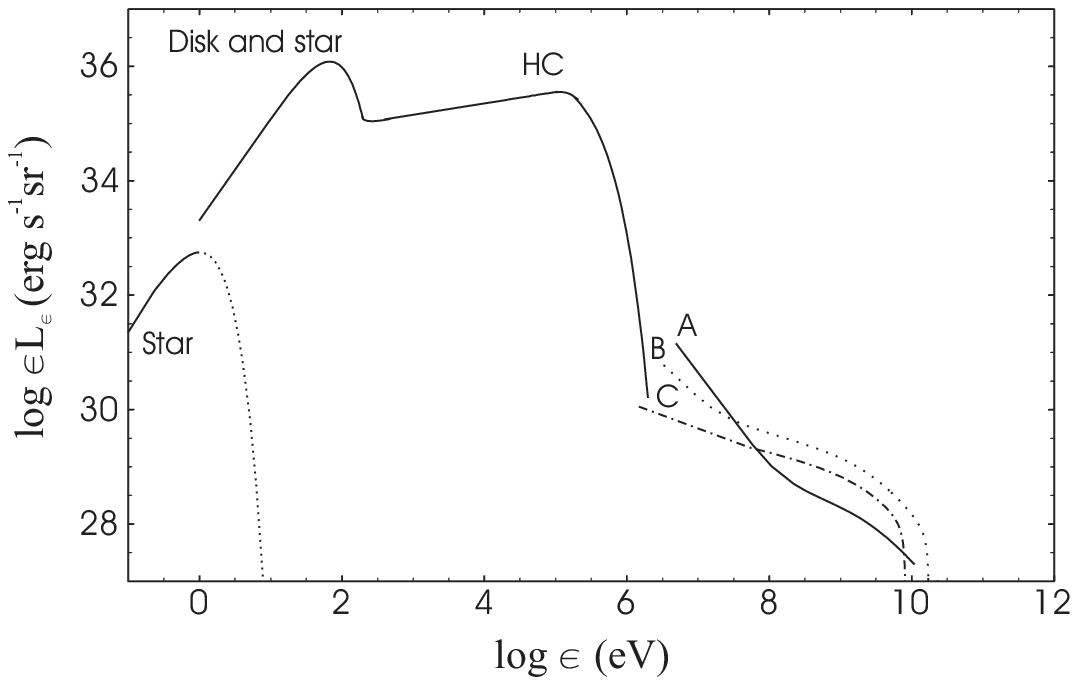} & (a)\\
\includegraphics[width=4in]{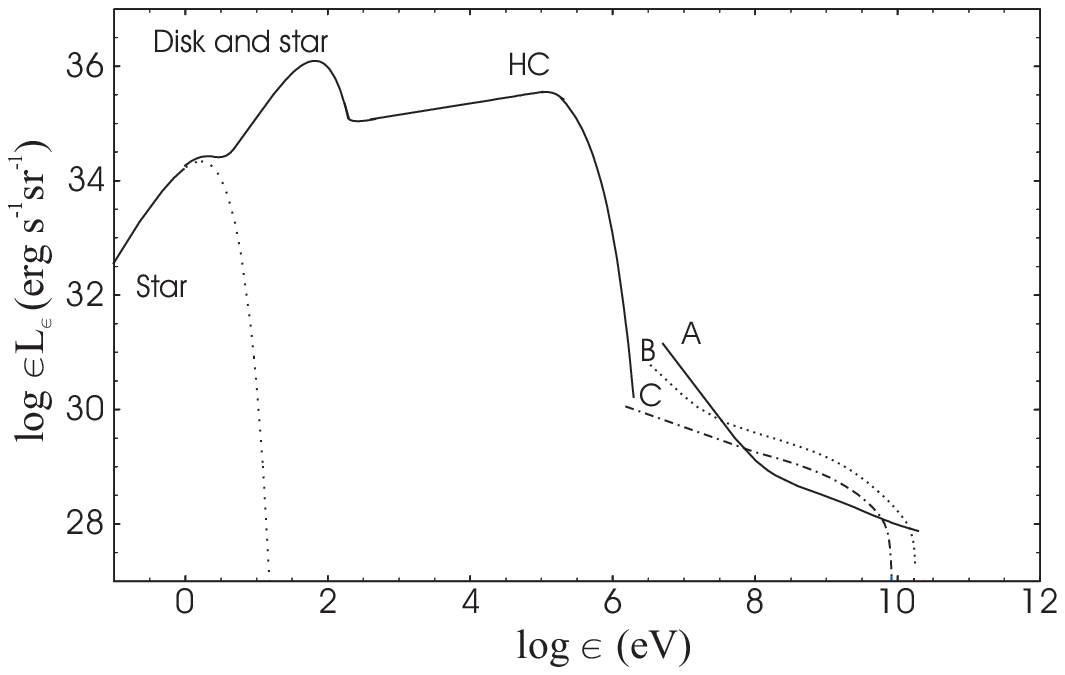} & (b)\\
\end{tabular}}
\caption{Spectral energy distribution of the EC emission from the jet of a microquasar with a K-M star (a) or
an F star (b) companion, seen at angles of $5^{\circ}$ (A), $15^{\circ}$ (B), and $30^{\circ}$ (C) from its
axis, for $\Gamma_{\rm jet}=3$, $q_{\rm jet}=1$\%, and an $E^{-2.3}$ pair spectrum between 1 MeV and 5
GeV.}\label{fig:탌soABC}
\end{figure}

XTE J1118+480 therefore serves as a good example for a low-mass microquasar and we adopt its characteristics
as input to our model (see table \ref{parset}). The spectral energy distributions (SED) of the thermal
stellar, disc and coronal components, taken from the afore mentioned publications, are displayed in Figure
\ref{fig:탌soABC}.a. The case of an F star companion is also considered in Figure \ref{fig:탌soABC}.b. We
calculate the EC emission from the three radiation fields assuming a population of $e^+-e^-$ pairs in a
persistent, cylindrical jet, with a power-law $E_e^{-p}$ distribution in number density per unit energy
between $E_{e\,\rm min}$ and $E_{e\,\rm max}$. The jet is assumed to be parallel to the disc axis, at an angle $\Phi$
to the line of sight. It moves with a bulk Lorentz factor $\Gamma_{\rm jet}$ and carries a total power $P_{\rm jet} =
q_{\rm jet} \dot{M}c^2$. The coronal emission is assumed to fill a sphere inscribed in the inner disc radius. The
particles are isotropically distributed in the jet frame and see the transformed energy densities from the
external seed photons. The cross-section for both the Thompson and Klein-Nishina regimes is used. The Compton
losses in the different regions modify the injected electron spectrum, introducing a break in the power-law
from an index $p$ to $p+1$ at the energy at which the cooling time equals the escape time. This is effective
in the disc radiation field. Another important ingredient is the absorption from two-photon pair creation in
the ambient radiation \cite{romero02}. It turns out to be quite effective near the disc while the coronal and
stellar fields are optically thin to the $\gamma$ rays. Back to the observer's frame, and to observed
energies shifted by the Doppler factor $D = \Gamma^{-1}_{\rm jet} (1-\beta_{\rm jet} \cos\Phi)^{-1}$, the stellar and
coronal contributions to the SED $\epsilon L_{\epsilon}$ are amplified by $D^{2+p}$ and the disc one by
$D^{2+p}(1-\cos\Phi)^{(1+p)/2}$ because of the anisotropic seed field \cite{dermer92,dermer02}. The former
amplification factor peaks along the jet axis and the latter near $\Phi = 15^{\circ}$ for $\Gamma_{\rm jet} \sim
3$ and $p = 2$ to 3. The jet synchrotron emission is only amplified by $D^{(3+p)/2}$.

This scenario differs from that of \inlinecite{georganopoulos02} who have imposed a much lower energy cutoff
to the electrons so that the hard X-ray emission result from their EC interactions with the stellar and disc
photons rather than from a hot inner accretion flow or from a hot coronal plasma energized by magnetic flares
above the disc.

Figure \ref{fig:탌soABC} shows the SED obtained per steradian in the laboratory frame, adding the
contributions from the three external photon fields. The coronal component (in the Klein-Nishina regime) is
preponderant in the COMPTEL 3-30 MeV band whereas the disc component (in the Thomson regime) takes over above
100 MeV. This is why the maximum $\gamma$-ray luminosity is reached for $\Phi$ close to 15$^{\circ}$,
reflecting the angle dependance of the disc amplification factor. The disc component is rather weak because
of the local absorption. It has a photon spectral index of 2.5 between 0.1 and 1 GeV. The stellar component
is small at all viewing angles, so the star spectral type has little impact on the $\gamma$-ray flux. These
results show that, even though the spectral index in the EGRET band matches that of the unidentified sources,
the maximum predicted luminosity, $L_{\rm max} \sim 4 \times 10^{29} (E/100\, \rm MeV)^{-0.5}$ erg s$^{-1}$ sr$^{-1}$, is 5 orders of magnitude too faint to account for the halo source flux at distances of 5 to 10 kpc.

%----------------------------------------------------------------
%----------------------------------------------------------------

%----------------------------------------------------------------
\begin{figure}
\centerline{\includegraphics[width=4in]{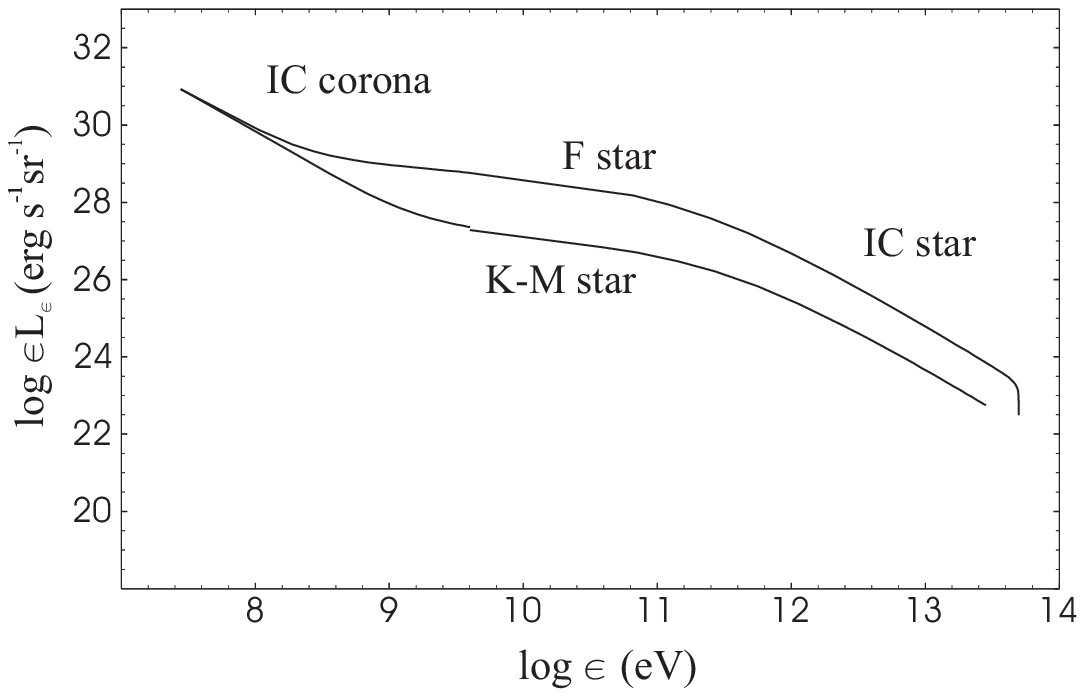}} \caption{Spectral energy distribution of
the EC emission from the jet of an extreme microblazar with an F or K-M star companion, seen at $1^{\circ}$
from its axis, for $\Gamma_{\rm jet}=10$, $q_{\rm jet}=1$\%, and an $E^{-2.3}$ pair spectrum between 5 MeV and 5
TeV.} \label{fig:킶lz}
\end{figure}
%----------------------------------------------------------------
The relative contributions of the three EC components change in the case of an extreme microblazar where the
bulk Lorentz factor $\Gamma_{\rm jet}$ reaches 10, electron energies extend to 5 TeV, and where the jet axis is
close to the line of sight ($\Phi = 1^{\circ}$). The generation of such a highly relativistic outflow, with
$\Gamma_{\rm jet} \geq 10$, has been recently observed from Circinus X-1, a neutron star with a stellar-mass
companion \cite{fender04}. Particle energies as high as 10 TeV have been inferred in the jets of the
stellar-mass microquasar XTE J1550-564, 0.1 pc away from the black hole, even though the jet is seen
deccelerating in the interstellar medium \cite{corbel02}. High-mass extreme microblazars are unlikely
counterparts to the low-latitude EGRET sources because of their extreme brightness. \inlinecite{bosch04}
predict a luminosity of $3 \times 10^{37}$ erg s$^{-1}$ sr$^{-1}$ at 1 GeV, well in excess of any EGRET source. Such
an event would appear as bright as Vela at 5 kpc. Figure \ref{fig:킶lz} shows the result for a low-mass
system. The soft EC emission from the corona predominates up to several hundred MeV for an F star and
to several GeV for a K-M star, beyond which the harder, $E^{-2.3}$, stellar component takes over. The disc EC
emission is negligible because of the lesser amplification at small aspect angle. The 35 times larger energy
density that the jet encounters around an F star compared with the K-M one results in a modest luminosity
increase by a factor $<$ 5 in the EGRET band. Anyhow, the very low luminosities, as well as the very soft
spectra, with photon indices ranging from 3 to 4, however, cannot account for the EGRET source
characteristics. Nor would these systems be detected above 100 GeV by the new-generation Cerenkov telescopes
since the predicted luminosity falls orders of magnitude below their sensitivity at a few kpc distance.

These objects can remain quiescent for years or decades before brightening by as much as $10^7$ in X rays in a
week. One should, however, keep in mind that the adopted keV luminosity of $\sim 10^{35}$ erg s$^{-1}$
sr$^{-1}$ corresponds to such a flaring state, i.e. to a bright X-ray source with a $\nu F_{\nu}$ flux of $2.6
\times 10^{-4}$ $(D/5\, {\rm kpc})^{-2}$ MeV cm$^{-2}$ s$^{-1}$. The GeV emission being mainly due to the disc or corona
(depending on the more or less extreme conditions in the jet), the predicted luminosity should be considered
as an upper limit for comparison with the unidentified sources. Therefore, EC emission in low-mass
microquasars largely fails to explain the variable EGRET sources at large scale heights.

%----------------------------------------------------------------
\begin{figure}
\centerline{\includegraphics[width=4in]{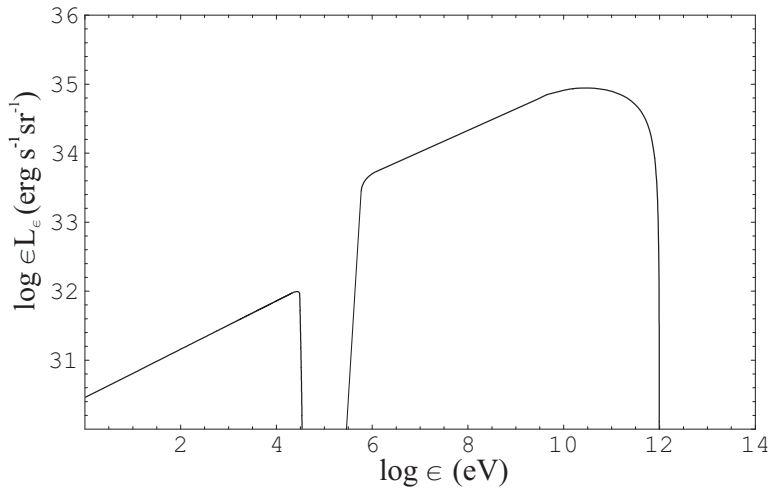}} \caption{Spectral energy
distribution of the synchrotron and SSC emission from the jet of an extreme microblazar, with a 10 G magnetic
field, seen at $1^{\circ}$ from its axis, for $\Gamma_{\rm jet}=10$, $q_{\rm jet}=0.1$\%, and an $E^{-2.3}$ pair
spectrum between 5 MeV and 50 GeV.} \label{fig:SSC}
\end{figure}
%----------------------------------------------------------------
In contrast with high-mass systems where the external radiation energy density largely surpasses the magnetic
one, SSC emission is likely to dominate even for a modest field strength of $\sim$ 10 G in the jet for which
the magnetic energy density compares with that of a K-M star. Applying the model developed by
\inlinecite{bosch04} to a cylindrical jet, one gets luminosities in excess of the necessary EGRET ones (see
Figure \ref{fig:SSC}). The synchrotron part of the spectrum is 100-1000 times fainter than the near-IR to UV
synchrotron emission recorded from XTE J1118+480 during the outburst, the adopted electron index in Figure
\ref{fig:SSC} being softer than that of XTE J1118+480 to avoid too hard spectra at high energy. The latter
softens beyond 10 GeV because of Klein-Nishina effects. One should keep in mind, however, that $\gamma\gamma$
absorption against the disc photons is not included although it efficiently limits the emerging $\gamma$-ray
flux. Adiabatic losses in an expanding jet would also modify the result although not severely. Nevertheless,
SSC emission in a low-mass microquasar appears as an interesting possibility to be investigated in greater
detail, bearing in mind that modelling the high-energy radiation from these fascinating objects is severely
limited by the highly uncertain choice of the jet bulk motion and magnetic field.

\acknowledgements
We thank Valenti Bosch-Ramon and J.M. Paredes for valuable discussions on leptonic models of MQs and details of some calculations. This work has been supported in part by an ECOS-Sud cooperation agreement between France and Argentina. GER also thanks additional support from Fundaci\'on Antorchas and ANPCyT.

\end{article}
\end{document}